\documentclass[pre,twocolumn,twoside,byrevtex,superscriptaddress,floatfix,longbibliography]{revtex4-1}

\usepackage[realmainfile]{currfile}
\usepackage{courier}
\usepackage{subfigure}
\usepackage{float}
\usepackage{graphicx}
\input{\currfilebase.settings}
\usepackage{color}

\usepackage{graphicx,epsfig,verbatim,enumerate}
\usepackage{amssymb,amsmath}
\usepackage{ifthen}

\usepackage{longtable}

\usepackage{mathtools}

\newboolean{twocolswitch}

\newcommand{\sindex}[1]{}
\newcommand{\nindex}[1]{}

\newcommand{\www}[1]{\url{#1}}

\usepackage{lettrine}

\setboolean{twocolswitch}{true}

\begin{document}

\title{\protect
Local information sources received the most attention from Puerto Ricans during the aftermath of Hurricane Mar\'ia

}

\author{
\firstname{Benjam\'in}
\surname{Freixas Emery}
}
\email{bfemery@uvm.edu}
\affiliation{
  Vermont Complex Systems Center,
  Computational Story Lab,
  The Vermont Advanced Computing Core,
  Department of Mathematics \& Statistics,
  The University of Vermont,
  Burlington, VT 05405, United States.
  }

\author{
\firstname{Meredith}
\surname{T. Niles}
}
\email{mniles@uvm.edu}
\affiliation{
  Department of Nutrition and Food Sciences, 
  The University of Vermont, 
  Burlington, VT 05405, United States
  }

\author{
\firstname{Christopher}
\surname{M. Danforth}
}
\email{cdanfort@uvm.edu}
\affiliation{
  Vermont Complex Systems Center,
  Computational Story Lab,
  The Vermont Advanced Computing Core,
  Department of Mathematics \& Statistics,
  The University of Vermont,
  Burlington, VT 05405, United States.
  }
  
\author{
\firstname{Peter}
\surname{Sheridan Dodds}
}
\email{pdodds@uvm.edu}

\affiliation{
  Vermont Complex Systems Center,
  Computational Story Lab,
  The Vermont Advanced Computing Core,
  Department of Mathematics \& Statistics,
  The University of Vermont,
  Burlington, VT 05405, United States.
  }


\date{\today}

\begin{abstract}
  \protect
  In September 2017, Hurricane Mar\'ia made landfall across the Caribbean region as a category 4 storm. In the aftermath, many residents of Puerto Rico were without power or clean running water for nearly a year. Using both English and Spanish tweets from September 16 to October 15 2017, we investigate discussion of Mar\'ia both on and off the island, constructing a proxy for the temporal network of communication between victims of the hurricane and others. We use information theoretic tools to compare the lexical divergence of different subgroups within the network. Lastly, we quantify temporal changes in user prominence throughout the event. We find at the global level that Spanish tweets more often contained messages of hope and a focus on those helping. At the local level, we find that information propagating among Puerto Ricans most often originated from sources local to the island, such as journalists and politicians. Critically, content from these accounts overshadows content from celebrities, global news networks, and the like for the large majority of the time period studied. Our findings reveal insight into ways social media campaigns could be deployed to disseminate relief information during similar events in the future.

\end{abstract}



\maketitle


\section{Introduction}
\label{sec:introduction}

\subsection{Background}
\label{sec:introBackground}

In the early morning of September 20th 2017, Puerto Rico was struck by its first Category 4 hurricane in 85 years \cite{meyer2017atlantic}. Hurricane Mar\'ia made landfall with wind speeds of 155 miles per hour, just 2 miles per hour short of qualifying as a Category 5 hurricane. The island's electrical grid, which had been left in highly vulnerable condition as a result of the territory's extreme debt \cite{allen2015npr}, was entirely wiped out. Most of the island was left without power for nearly two months, and some areas remained without power for nearly a year \cite{sullivanEmily2018npr}.

The Puerto Rican government accepts an independent study approximating that 2,975 deaths resulted from the storm and its aftermath \cite{gwu2018},  although estimates from other researchers place 8,000 excess deaths within the range of plausible realities\cite{doi:10.1056/NEJMsa1803972}. An internal report at the Federal Emergency Management Agency (FEMA) acknowledges a failure to adequately prepare for the 2017 hurricane season, resulting in the agency being unable to properly support the victims of Hurricane Mar\'ia \cite{sullivan2018npr}. Failures at the local level also played a role in the resulting damage. In July 2019, two Puerto Rican officials were arrested on corruption charges for allocating \$15.5 million in relief contracts to favored businesses, despite credible accusations of being unqualified \cite{allyn2019npr}. These arrests occurred among a series of events that brought hundreds of thousands of Puerto Ricans to protest in the streets for two weeks in demonstrations that eventually led to the resignation of Puerto Rican governor Ricardo Roselló \cite{ocasio2019governor}.

Despite 89\% of cell phone towers being offline \cite{fitzgerald2017cellPhones}, many Puerto Ricans were connecting with each other in the first few days after the storm. Individuals and families were able to find isolated hot spots of cellphone coverage, and despite only being able to communicate with one another in person or with battery-operated radios, knowledge of these spots spread rapidly. During the first week after the storm, these spots could be found with dozens of cars pulled over, their passengers outside holding smartphones in the air \cite{mazzei2017cellPhones}.

At the time, Hurricane Mar\'ia was the latest in a sequence of natural crisis events spanning nearly a decade in which social media was used for practical communication. Despite causing the highest death toll in 20 years and the largest cumulative blackout in United States history \cite{irfan2018blackout}, Hurricane Mar\'ia received little attention from the mainstream media \cite{chang2018ignored}. Our goal in the present work is to retrospectively investigate and reveal specific details about the nature of public discussions about Hurricane Mar\'ia using tools from network science and information theory.

\subsection{Existing research}
\label{sec:introExistingResearch}

Work surrounding previous major natural hazard events in the contiguous United States has found that the temporal distribution of disaster-related tweeting varies between disaster types, and for hurricanes, the greatest volume of tweeting tends to happen in the anticipation of the event\cite{niles2018average}. Additionally, users who demonstrate the greatest increase in activity during the timeline of such an event tend to be those with an average-sized following, rather than highly connected individuals. There has also been work showing that people are largely motivated to spread information on Twitter during disaster events by the belief that the information is important and believable \cite{abdullah2017retweet}. Researchers have also found that during hurricanes Twitter can be used to predict how to best focus recovery efforts \cite{guan2014using}, and that Twitter activity correlates with hurricane damage \cite{kryvasheyeu2016rapid}. Algorithms have been developed to identify flood victims asking for help on Twitter \cite{singh2017event}. Going beyond publicly available social media, research have found that private SMS messaging groups have been integral to the spread of information among small communities during hurricanes \cite{chu2019building}.

Twitter in particular lends itself well to a network framework for analysis. Although Twitter does not make follower and following lists easily accessible, public messages between accounts can be easily identified. Research into protest movements has used such interactions between accounts to specify a network representation \cite{jackson2015hijacking,jackson2016ferguson,jackson2019women}. Other work has analyzed the network topology of hashtag topics and used information theory to identify the differences in discourse between hashtag topics \cite{gallagher2018divergent}. The use of network analysis on Twitter surrounding disasters is not new. For example, after the Deepwater Horizon oil spill, Twitter networks were used to study the online conversation as the recovery efforts unfolded \cite{sutton2013tweeting}.

\subsection{Focus of study}
\label{sec:introFocus}
We wish to understand the online conversation on Twitter surrounding Hurricane Mar\'ia, beginning more broadly and then narrowing down to those directly affected. We begin by looking specifically at the differences between English and Spanish tweets.

Following our more general analysis, we narrow our focus to those on the island or with close connections to people on the island. We seek to identify how Twitter could potentially have been leveraged to increase the effectiveness of aid to Puerto Rico during Hurricane Mar\'ia. If public officials want to inform the population about hurricane risk, who are the best people on Twitter to provide information? Does the answer change when the topic shifts from warning to aid? Do they change over time as the cycle from anticipation to aftermath unfolds?

\section{Methods}
\label{sec:methods}
\subsection{Data collection}\label{sec:methodsData}
\subsubsection{Between-language analysis}\label{sec:methodsLanguageData}
We used Twitter's Decahose sample, a paid-access API providing a pseudo-random 10\% sample of all public tweets. We sought to collect the conversations on Twitter surrounding the hurricane and Puerto Rico around the time that the island was hit. Over the course of 30 days, from September 16 to October 15 in 2017 (inclusive), we collected all tweets that contained the strings ``puerto", ``rico", ``hurricane", ``maria", ``huracan", ``mar\'ia", or ``hurac\'an". We used Twitter's provided language tag to bin the Tweets into Spanish and English corpora. We record the author of each tweet, it's unique tweet identification number, the date and time of it's publication, and the content of the tweet. Our resulting collection contains 1,947,489 Spanish tweets and 3,030,286 English tweets.

\begin{figure}[tp!]
\centering
\includegraphics[width=\columnwidth]{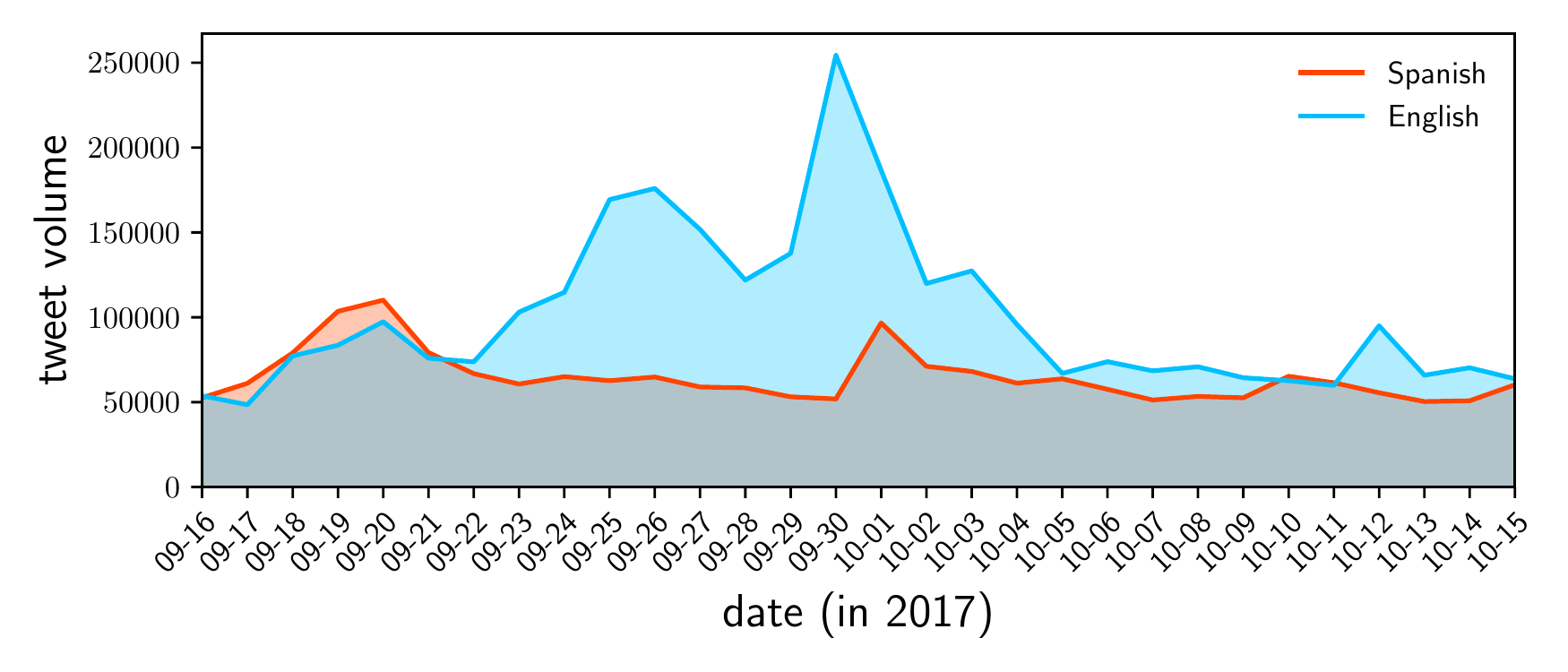}
\caption{The daily volume of Spanish and English tweets concerning the hurricane or Puerto Rico in our collected subset of Twitter's Decahose sample, roughly 10\% of all public tweets.}
\label{fig:volume}
\end{figure}

\subsubsection{Hashtag analysis}\label{sec:methodsDataHashtag}
Using the same data source, we gathered all Decahose tweets containing at least two hashtags and constructed a list of co-occurring hashtag pairs on each day: a hashtag topic network. We then extracted the subset of the network that only includes hashtags co-occurring with \#PuertoRico. This network subset is also known as the ``ego network'' of \#PuertoRico. The resulting timeline of networks contains 11,633 tweets and 7,392 unique hashtags.

\subsubsection{Victim proxy}\label{sec:methodsDataVictim}
Our goal here was to capture a set of tweets that could serve as a proxy for the conversation among residents of Puerto Rico and people with personal connections to those residents. We began by finding the user names for accounts located on mainland Puerto Rico, or any of the smaller islands of Vieques, Culebra, or Mona during the hurricane. Looking through all geolocated tweets, we saved account names for those who authored tweets from the island on September 20 or 21, 2017. A total of 93 seed users sent geolocated tweets from Puerto Rico during the hurricane. Using Twitter's free Premium API, we searched for all tweets that mentioned or retweeted a user from this seed list during the date range specified in Section \ref{sec:methodsLanguageData}. This extended our list of account names to 1,511 users, hereafter our ``user sample".

We chose our time interval to maximize preparatory time while avoiding overlap with the recovery from Hurricane Irma, which had hit only days before. This choice of preparatory time interval is also consistent with findings by Niles et al., which demonstrate it to be valid for showing tweet increases during a climate emergency \cite{niles2018average}. We extended the studied interval for a month, in order to look at the change in communication behavior throughout the development of the situation and the revelations of new information. This allowed us to gather additional users who were on the island but did not enable geolocation, as well as people elsewhere who are personally connected to people on the island during the hurricane. 

With our user sample constructed, we recorded every Decahose tweet that (1) was authored by an account on our list, (2) was marked by Twitter as a Spanish tweet, and (3) either mentioned or retweeted another user. We also restricted our corpus to Spanish tweets, as information that's truly targeted at the Puerto Rican populations will be in Spanish, the island's primary language. From our list of tweets, we kept only those using hand-selected vocabulary that indicated a probable discussion of the hurricane's effects or need for supplies, shown in Table \ref{table:keyWords}. These words are adapted from the list of words mentioned in Section \ref{sec:introExistingResearch}, with the addition of words ``puerto rico'' and ``hurac\'an'' and ``mar\'ia''.

\begin{table}[h!]

\tiny
\centering
\begin{tabular}{l|lllll}
\toprule
Spanish &               agua &       bebida &  generador &        maria &   suministros \\
        &           albergue &        comida &    huracan &      mar\'ia &  supermercado \\
        &           alimento &     corriente &  hurac\'an &       prepar &        tienda \\
        &         asistencia &  desprevenido &      inund &  puerto rico &         toldo \\
        &              ayuda &   en conserva &       lluv &    refrigera &        viento \\
        & banco de alimentos &      enlatado &        luz &      refugio &               \\
\hline
English &              water &         drink &  generator &        maria &   supplies \\
        &             hostel &        food   &   hurricane &     mar\'ia &  supermarket \\
        &               food &     current   &   hurricane &     prepare &        store \\
        &         assistance &    unprepared &      flood &  puerto rico &         tarp \\
        &              help  & nonperishable &       rain &   refrigerat &        wind \\
        &           foodbank &        canned &      power &       refuge &               \\
\hline
\end{tabular}
\caption{Words used to flag tweets in our study (top) and their English translations (bottom).}

    \label{table:keyWords}
\end{table}

\subsection{Hedonometric analysis}
For their 2015 work, Dodds et al. collected ``happiness'' ratings of 10,000 words across ten languages \cite{dodds2015human}. Words were rated by survey participants on a 9-point, double-Likert scale from 1 (most negative) to 9 (most positive), with 5 representing a neutral word. The resulting collection of word lists and continuous sentiment scores is known as the language assessment by Mechanical Turk or ``LabMT'' dataset. For these analyses, we assign words their score identified by Dodds et al.  In order to improve our ease of analysis, we subtract 5 from the score of each word, giving unhappy words a negative score. With this scoring scheme negative words in Spanish like ``odiar" (``to hate'') and ``guerra'' (``war'') respectively have scores of -2.92 and -3.22, where as positive words like ``amar'' (``to love'') and ``beso'' (``kiss'') have scores of 3.38 and 3.36. Similarly for English, ``joy'' has a score of 3.16 and ``terror'' has a score of -3.24. We compute the sentiment of either language for some time period as
\begin{equation}
    \label{eq:sentiment}
    \langle h \rangle = \sum_i h_i p_i,
\end{equation}
where $h_i$ is the sentiment score of a particular word, and $p_i$ is that word's relative frequency.

\begin{figure}[tp!]
\centering
\includegraphics[width=\columnwidth]{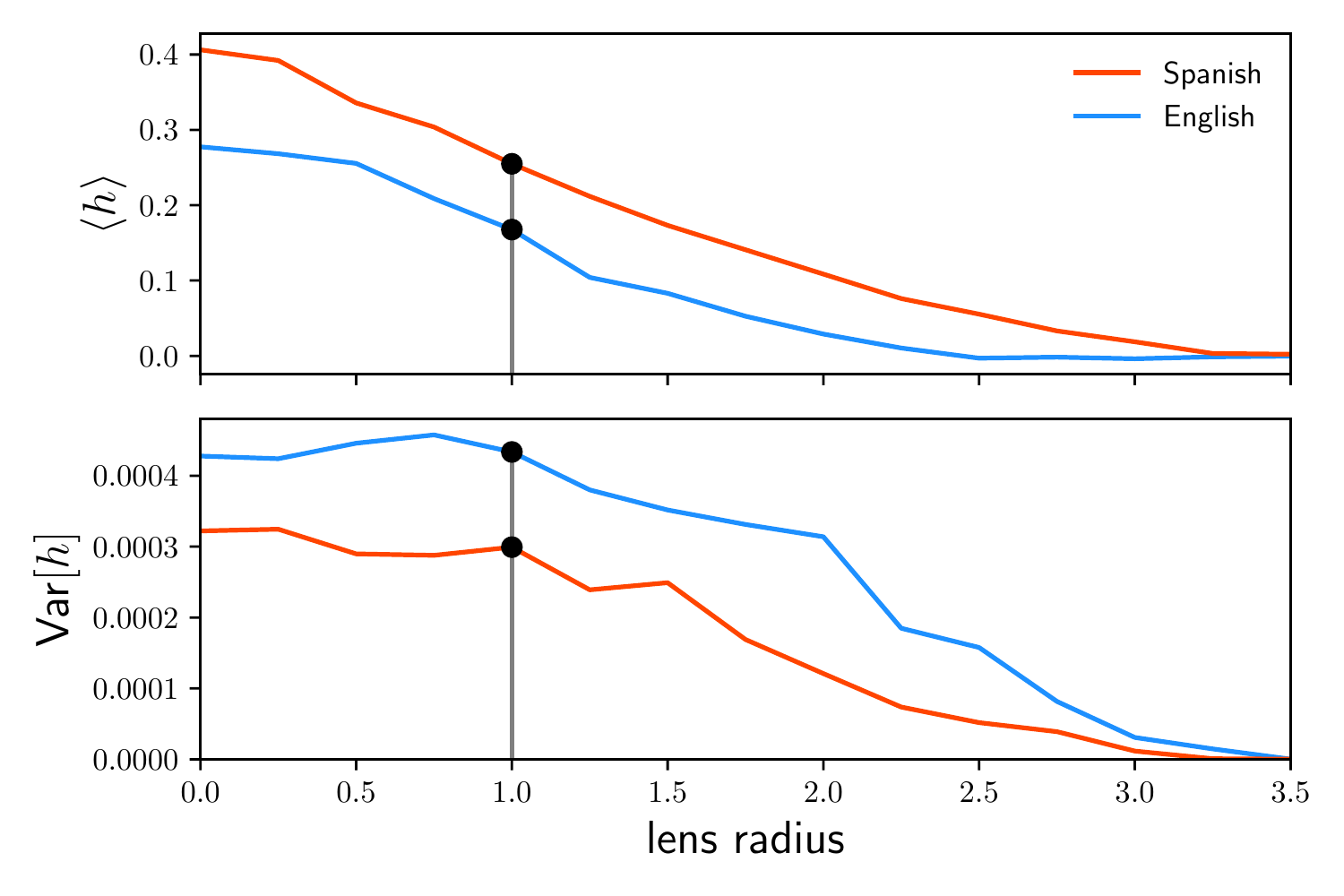}
\caption{The mean and variance of daily sentiments at each lens radius value for English and Spanish. At all lense values, Spanish has a higher mean happiness score than English, but a smaller spread. We choose out lens size to remove a sufficient amount of neutral words, which in mass can skew the scores and make results less intuitive. A gray vertical line in each plot indicates the selected lens radius. Black dots show the values of the mean and variance of the daily sentiments at the selected lens radius. This value is selected because it is roughly the edge of the variance's plateau for the two languages (see \cite{dodds2011temporal} for further support).}
\label{fig:lensSweep}
\end{figure}

For meta-analysis of the happiness of a corpus, we wish to remove words with scores within some radius of zero (neutral). Words with scores close to neutral are removed from consideration to increase the strength of the emotional signal recovered from the corpus. These masked words are largely comprised of either (a) emotionally neutral words or (b) context dependent. Attempts at sentiment analysis prior to the creation of LabMT involved the removal of hand-selected ``function'' words, but it has been found that many words that individuals might label as ``function'' have a substantial bias in sentiment \cite{dodds2015human,garcia2015language}. One can therefore deduce that it's more accurate to remove words that have been empirically determined to have near-neutral net sentiment \cite{dodds2011temporal}.

To determine the appropriate width for this sentiment mask, we sweep through values and compute the mean and variance of the sentiment scores of all tweets for English and Spanish from Section \label{sec:methodsData} (Fig. \ref{fig:lensSweep}). While the mean for both languages monotonically decreases with increasing neutral mask, the variance has a plateau from 0 to 1, after which it drops. For this reason, we choose a lens radius of 1, which is also consistent with previous work using the labMT dataset \cite{dodds2011temporal,reece2017forecasting,cody2016public,reagan2016emotional,frank2013happiness,mitchell2013geography}. 

\begin{figure*}[tp!]
\centering
\includegraphics[width=\textwidth]{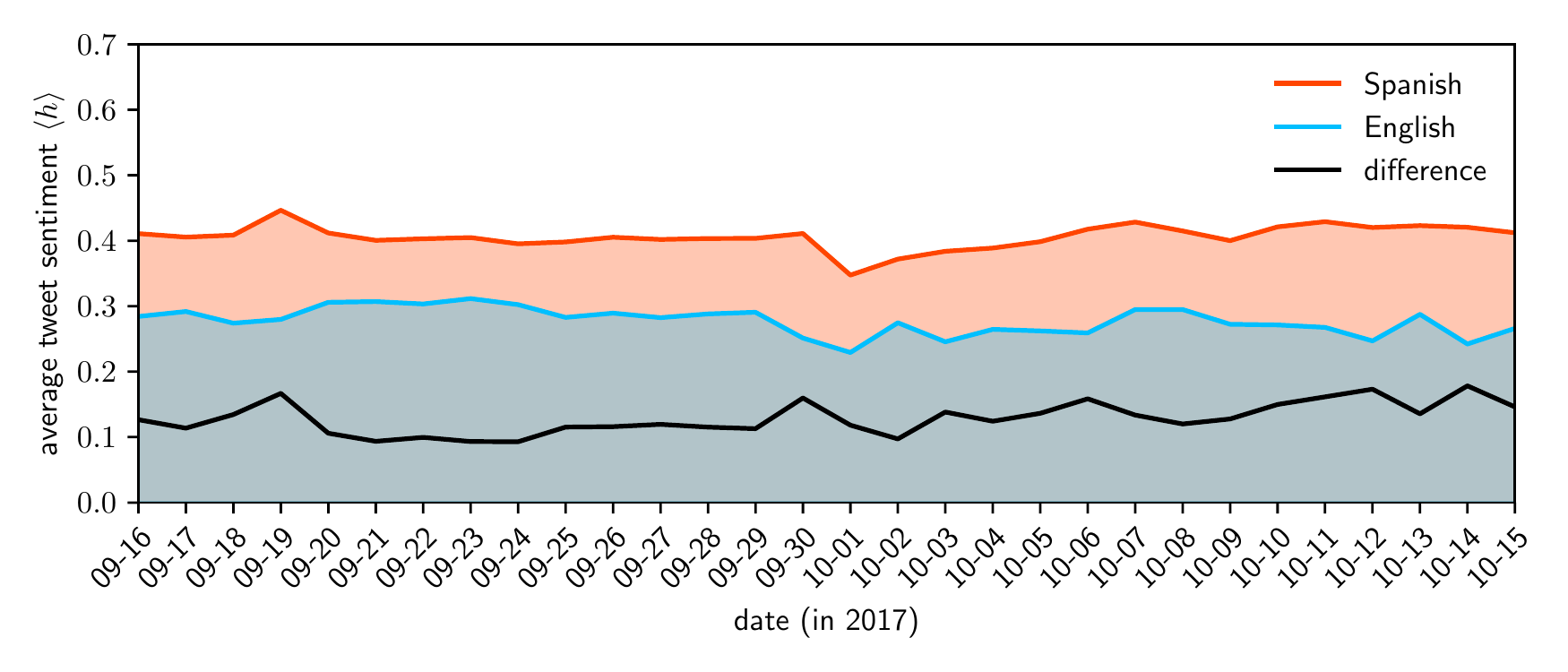}
\caption{The daily average happiness of Spanish and English tweets in our collected subset of Twitter's Decahose sample. The difference between the average happiness for each day is shown by the solid black line. Spanish tweets are more positive than English tweets throughout the time period, likely due to the underlying word scores. In previous work studying 10 major languages Dodds et al.\ found Spanish to be the language containing the highest proportion of happy words \cite{dodds2015human}.}
\label{fig:happs}
\end{figure*}

\subsection{Network specification}

\subsubsection{Hashtags}\label{sec:methodsNetworkHashtag}

We generate an aggregate network by superimposing the daily ego networks described in Section \ref{sec:methodsDataHashtag}. We remove the ego node, \#PuertoRico, which results in many nodes and small components of nodes becoming disconnected from the largest component of the graph. We remove all components apart from the largest connected component, and conduct the remainder of the analysis on this component of the network. Using the Louvain clustering algorithm, we identify five communities within the network structure \cite{blondel2008fast}.

\subsubsection{Users}\label{sec:methodsNetwork}

In order to interpret the flow of information among the individuals of interest, and identify the distribution of influence, we generate a network of Twitter users connected by retweets and mentions. Each tweet pulled from the Decahose as described in Section \ref{sec:methodsDataVictim} constitutes a directed link from the author to the mentioned or retweeted user. 

We identify communities in the network topology with the Louvain clustering algorithm. Inspection of the network's tweet content revealed that the second largest community consisted almost entirely of people involved in the 2017 Venezuelan protests, and several smaller communities shared this property \cite{wikiVzla}. To modify the network to better represent the focus of the present study, we counted the proportion of tweets in the largest Venezuelan network that contained the words ``venezuela'', ``vzla'', or ``maduro'', and computed the measurement's Wilson interval lower bound at 99\% confidence \cite{wilson1927probable}. We then removed all communities with a proportion of tweets containing those same words at least as high as that lower bound.

The resulting network has 2,011 users (nodes) connected by 2,466 tweets (edges). Although our links are generated from a 10\% sample of tweets and we reduce our corpus by content type, we maintain a third of the nodes from our original user list due to the heavy-tailed nature of online social networks \cite{kumar2010}. Finally, we hand-identified the fifty largest nodes by in-degree as being news outlets, politicians, citizens, weather stations, meteorologists, or journalists.

We consider three major attributes of the aggregate network: the in-degree distribution, density, and the average shortest path length. The in-degree of a node is the number of links connecting other nodes to it (tweets mentioning or retweeting the corresponding user). We define the density of the network as
\begin{equation}
    \label{eq:density}
    d = \frac{m}{n(n-1)},
\end{equation}
where $n$ is the number of nodes and $m$ is the number of directed links in the network, each counted once regardless of weight. This is in essence the fraction of possible locations for a link where there exists one. We compute the average shortest pathlength by iterating over each ordered pair of nodes $s$ and $t$ and finding the minimum number of links defining a path from $s$ to $t$. From these pathlengths, we compute the average shortest pathlength as \begin{equation}
    \label{eq:avgPathLength}
    a = \sum_{s,t\in V}\frac{p(s,t)}{n(n-1)},
\end{equation}
where $V$ is the set of nodes, and $p(s,t)$ is length of the shortest path from node $s$ to node $t$.

\subsection{Shannon entropy}\label{sec:methodsEntropy}

We rely on information theoretic measurements in our analysis of tweet content. The most basic measurement we use is Shannon's entropy $H$, which describes the diversity of the probability mass function of nominal variables. Shannon's entropy of a corpus with $n$ unique words is defined as
\begin{equation}
    \label{eq:H}
    H = -\sum_{i=1}^{n}p_{i}\log_{2}p_{i},
\end{equation}
where the $i$th word appears with probability (or rate) $p_{i}$. Shannon entropy is maximized when every word occurs with equal probability, and approaches 0 as the corpus becomes dominated by one word. 

\subsection{Jensen-Shannon divergence}\label{sec:methodsJSD}
To compare word distributions for two subsets of our corpus, we compute the Jensen-Shannon divergence between them. 

If we consider any two corpora, $T_1$ and $T_2$, with word frequency distributions $P_1$ and $P_2$, and their combined distribution $M=\pi_{1}P_1 +\pi_{2}P_2$, where $\pi_1$ and $\pi_1$ are the relative sizes of the two corpora such that $\pi_1+\pi_2=1$, the Jensen-Shannon Divergence $D$ is given by
\begin{equation}\label{eq:JSD1}
    D(P_1||P_2) = \pi_1\sum_{i=1}^{k}p_1,i\log_{2}\dfrac{p_{1,i}}{m_{i}} + \pi_2\sum_{i=1}^{k}p_1,i\log_{2}\dfrac{p_{1,i}}{m_{i}}.
\end{equation}
This expression is the weighted average of each corpus's Kullback-Leibler divergence from the mixed corpus.

We use the linearity of Jensen Shannon Divergence extensively when comparing corpora in order to identify the words and topics that contribute most to their difference.

\begin{figure}[tp!]
\centering
\includegraphics[width=.45\textwidth]{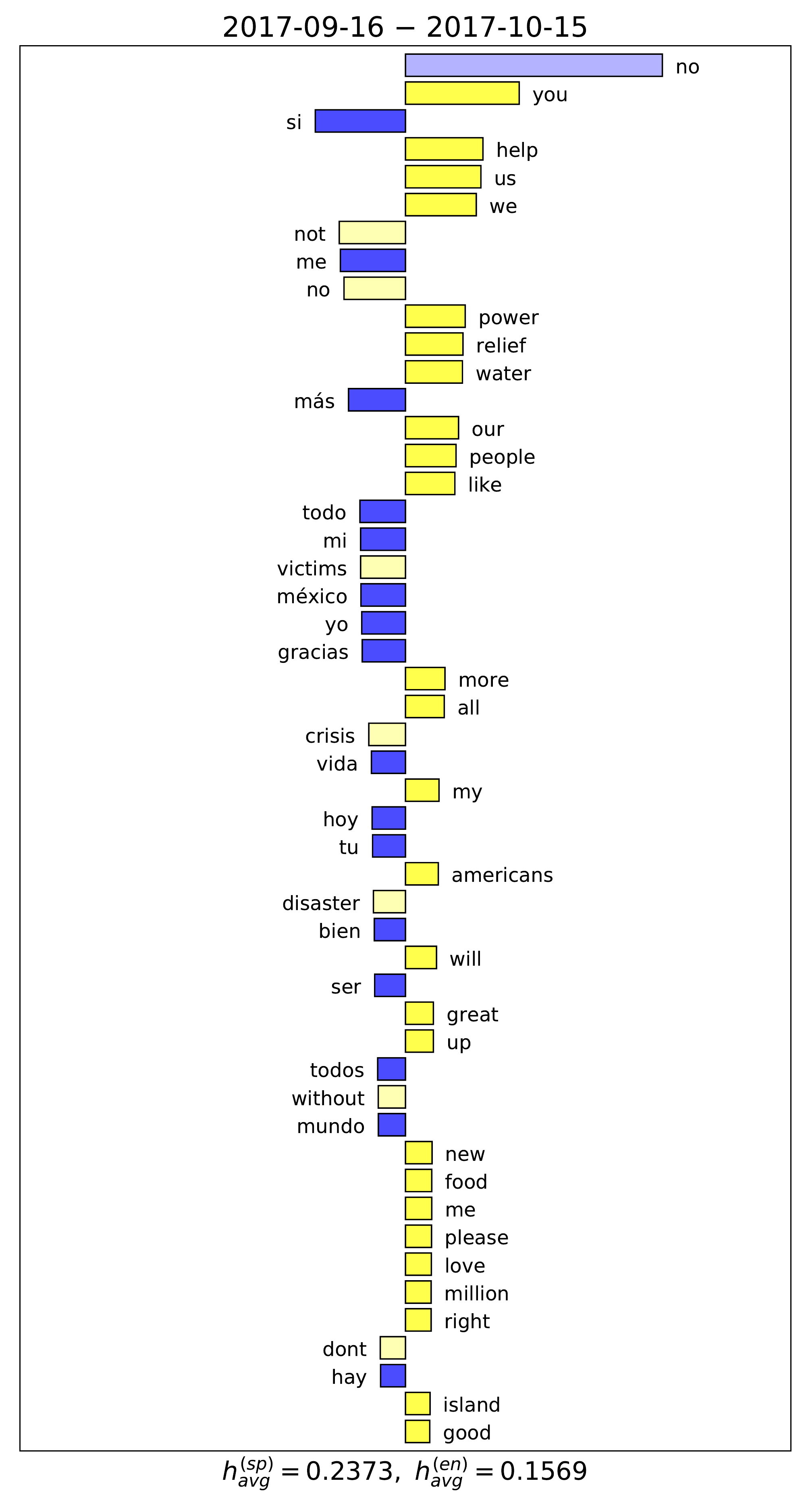}
\caption{Sentiment wordshift for the entire studied time span, 2017-09-16 through 2017-10-15. Each bar indicates the contribution of a word to the difference between the average happiness of Spanish tweets vs English tweets. The length of the bar is the product of the word's relative sentiment score and its relative frequency. We consider consider words to be ``happy'' if they have a LabMT score greater than 5 (positive adjusted score), and ``sad'' if their LabMT score is less than 5 (negative adjusted score). Bars that extend leftward are either happy Spanish words (deep blue) or sad English words (muted yellow), and bars that extend to the right are either happy English words (bright yellow) or sad Spanish words (muted blue).}
\label{fig:wsAll}
\end{figure}

\begin{figure}[tp!]
\centering
\includegraphics[width=.45\textwidth]{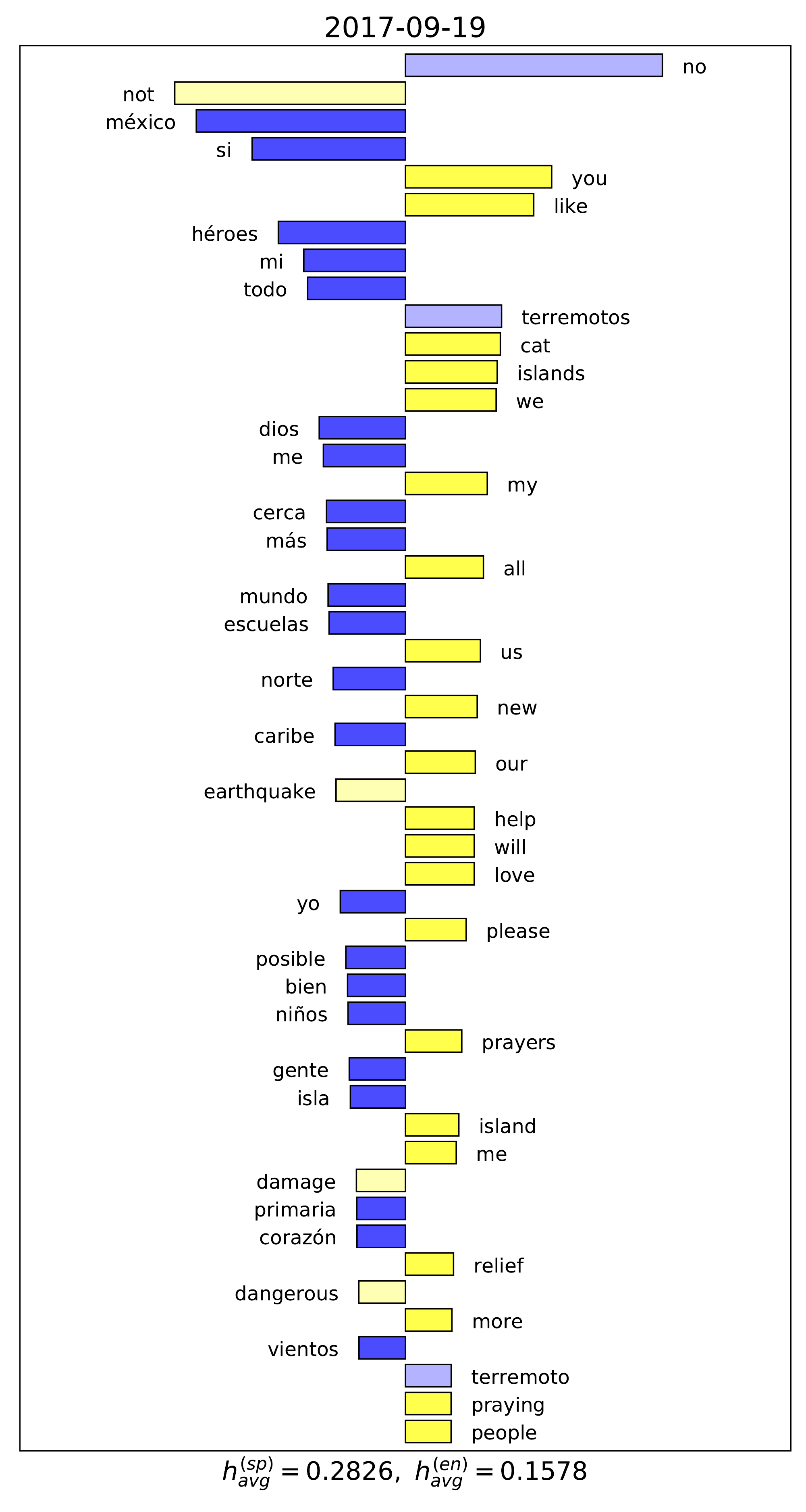}
\caption{Sentiment wordshift for the individual day on 2017-09-19. See Figure \ref{fig:wsAll} for details on the methodology.}
\label{fig:wsEarlyLate}
\end{figure}

\section{Results}
\label{sec:results}
\subsection{Daily volume and happiness}

We show the time series for daily tweet volume for each language is shown in Fig. \ref{fig:volume}. Both the Spanish and English tweet volume rises in anticipation of the hurricane, reaching a maximum on the date of landfall. The English volume begins steadily rising again two days later, eventually reaching another peak on September 26th. A sudden spike appears in the English tweet volume on September 30, following a string of tweets from Twitter user @realDonaldTrump, blaming the mayor of San Juan for poor leadership. After this spike, English tweet volume decays to a steady 50,000 per day. Spanish Tweet volume sees a much smaller but still prominent spike on October 1. 

In Fig. \ref{fig:happs}, we show the daily average happiness of Spanish and English tweets, as well as their difference on each day. Around the date of landfall, English sentiment remains unchanged, while Spanish sentiment spikes upward. We also see downward spikes in the sentiments of both languages on October 1. English also sees spikes on October 2 and October 13 that are unmatched in the Spanish corpus.

\subsection{Between-language wordshifts}

From here, we examine the contribution of individual words to the difference in average happiness between the two languages over periods in time. For this we use a visualization tool called the ``wordshift''. Wordshifts for different dates and date ranges are shown in Figure \ref{fig:wsAll}. These wordshifts are distinct from wordshifts used in previous work, because the corpora being compared must be treated as having a completely unique sets of words. Because of this, there is no sense in comparing whether or not a certain word is less or more frequent in a certain corpus. Instead, we look at which words in each corpus contribute the most to bringing the happiness of that corpus up or down. Words on the left of the shift are thus making Spanish look like the happier language, and words on the right make English appear happier. We provide more information on this computation in the caption of Fig. \ref{fig:wsAll}.

\begin{figure*}[tp!]
\label{fig:network}
\centering
\includegraphics[width=.8\textwidth]{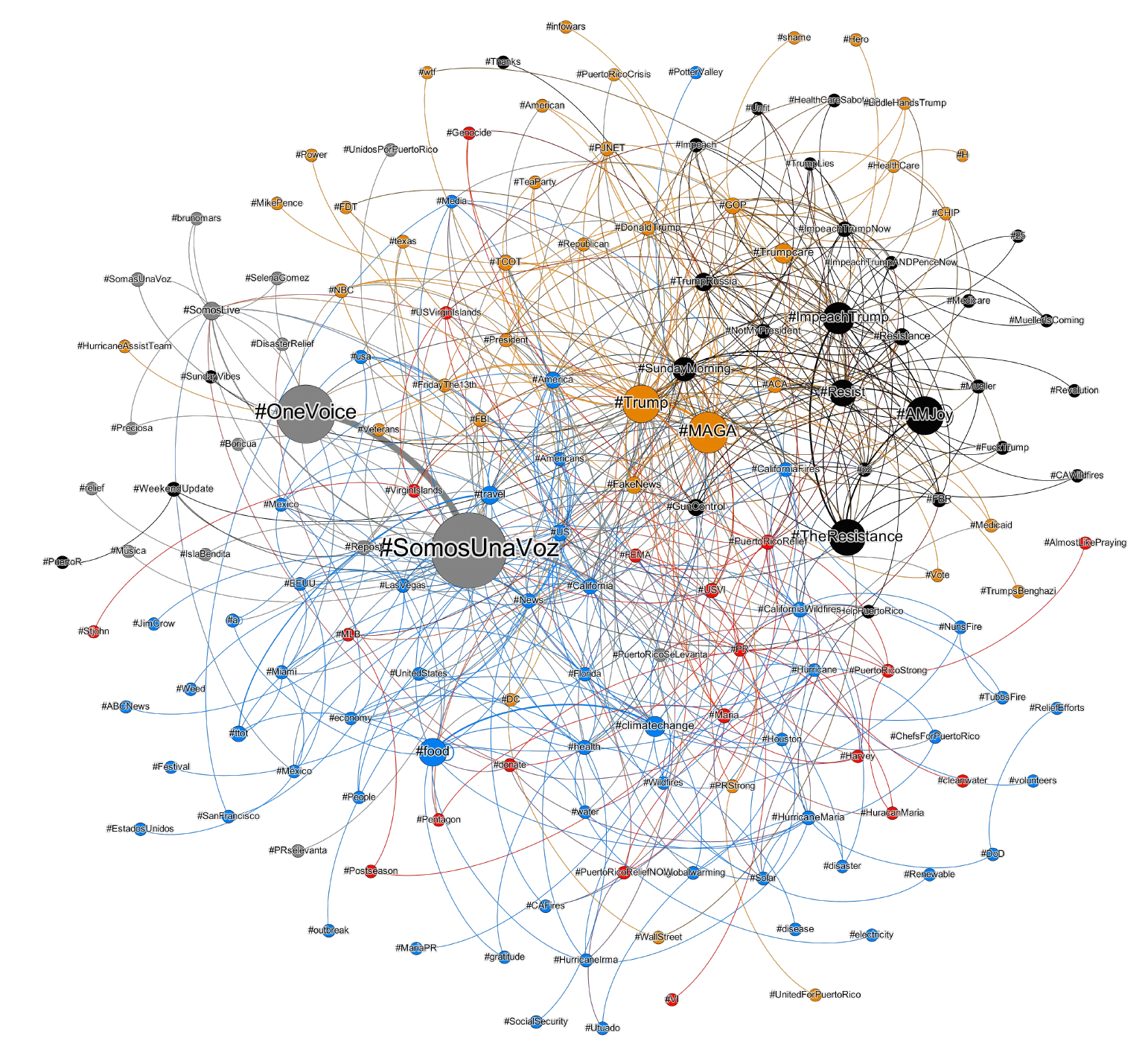}
\caption{The superposition of daily hashtag ego networks for \#PuertoRico. Nodes are sized proportionally to the degree of each node, and edges are resized proportionally to their weight, a representation of the number of co-occurrences of the hashtags at either end of the edge.}
\label{fig:hashtagNetwork}
\end{figure*}

In Fig. \ref{fig:wsAll}, we present the sentiment wordshift between Spanish and English for the entire studied time interval. We find that both sides are mostly occupied by ``happy'' words of each language, and the ``sad'' words that contribute are almost all negations, such as the Spanish ``no'' and the English ``no'', ``not'', and ``don't'', all of which translate to ``no'' in Spanish. Other prominent negative words in the wordshift are ``without'' and ``lost''.

It is also notable that most of the ``happy'' words contributing to the wordshift are likely being used in a negative context. For example, ``power'', ``relief'', and ``help'' all in part refer to services the island has lost or needed. The word ``m\'exico'' appears prominently for two reasons. First, the LabMT word list was evaluated by Mexican Spanish speakers, so it is reasonable to suspect ``m\'exico'' to receive a higher score than if it had been evaluated by a more globally representative sample of Spanish speakers. Second, an earthquake occurred in Mexico around the same time, prompting people of the Latinx community to talk about the disasters happening to two major geographic parts of Latin America.

Fig. \ref{fig:wsEarlyLate} highlights the words contributing to the difference in happiness between English and Spanish tweets on 2017-09-19. We choose this date because it is the day before Hurricane Mar\'ia made landfall, and because we observe a spike in the sentiment difference between Spanish and English on this date. The words that appear indicate emphasis on the earthquake in Mexico. Some of the words accounting for the upward spike in tweet sentiment at on that day are words such as ``h\'eroes'' (``heroes'') and ``dios'' (``God''), which focus on positive aspects of the events.



\subsection{Hashtag network}
\label{sec:resultsHashtagNetwork}

The network of hashtags for the studied date range consists mostly of small nodes with in-degrees in the single-digits (see Fig. \ref{fig:hashtagNetwork}). A small handful of larger nodes dominate, which exist in each of the detected communities. Notable in the network are two closely connected communities containing content about U.S. President Donald Trump. One community (indicated with black nodes) contains slogans attacking Trump such as \#Resist, \#Resistance, and \#ImpeachTrump, as well \#AMJoy, a hashtag referring to MSNBC reporter Joy Reid's news show (Reid visited Puerto Rico on October 14th). The other community contains topics in support of the president, such as \#Trump, \#MAGA, and \#FakeNews, but also contains anti-Trump topics, indicating that some hijacking of those pro-Trump hashtags took place. A third community is dominated by the hashtags \#SomosUnaVoz and \#OneVoice, both promotional hashtags for a live televised charity concert for disaster relief, which occurred on October 14th as well. 

Two further communities, denoted with blue and red nodes, contain hashtags indicating discussion about the hurricane and other recent disaster events, as well as their respective relief efforts. 

As a whole the hashtag network demonstrates a large variety of conversation topics, but a generally disproportionate emphasis on Donald Trump that dwarfs conversation about the storm, relief efforts, and other directly related topics.

\subsection{Communication network}
\label{sec:resultsNetwork}

\begin{figure}[tp!]
\centering
\includegraphics[width=\columnwidth]{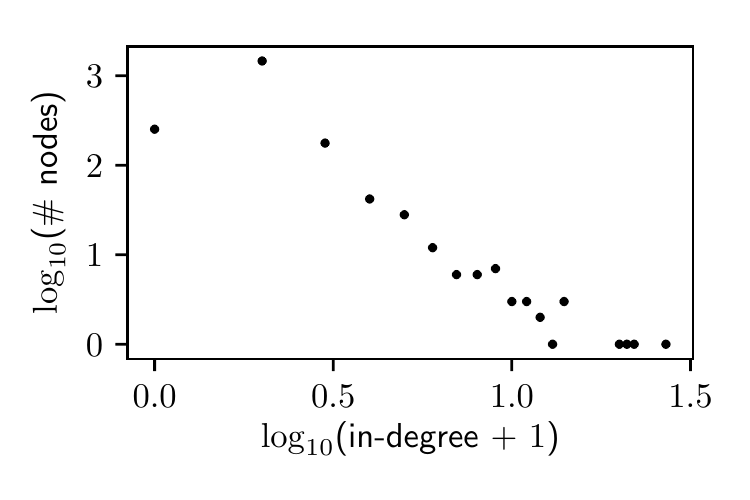} \caption{
    The in-degree distribution of the communication network (shown in Fig \ref{fig:networka}), plotted on a log-log scale.
  }
\label{fig:degreeDist}
\end{figure}

\begin{figure*}[tp!]
\label{fig:network}
\centering
\subfigure[\ Total]{\label{fig:networka}\includegraphics[width=.8\textwidth]{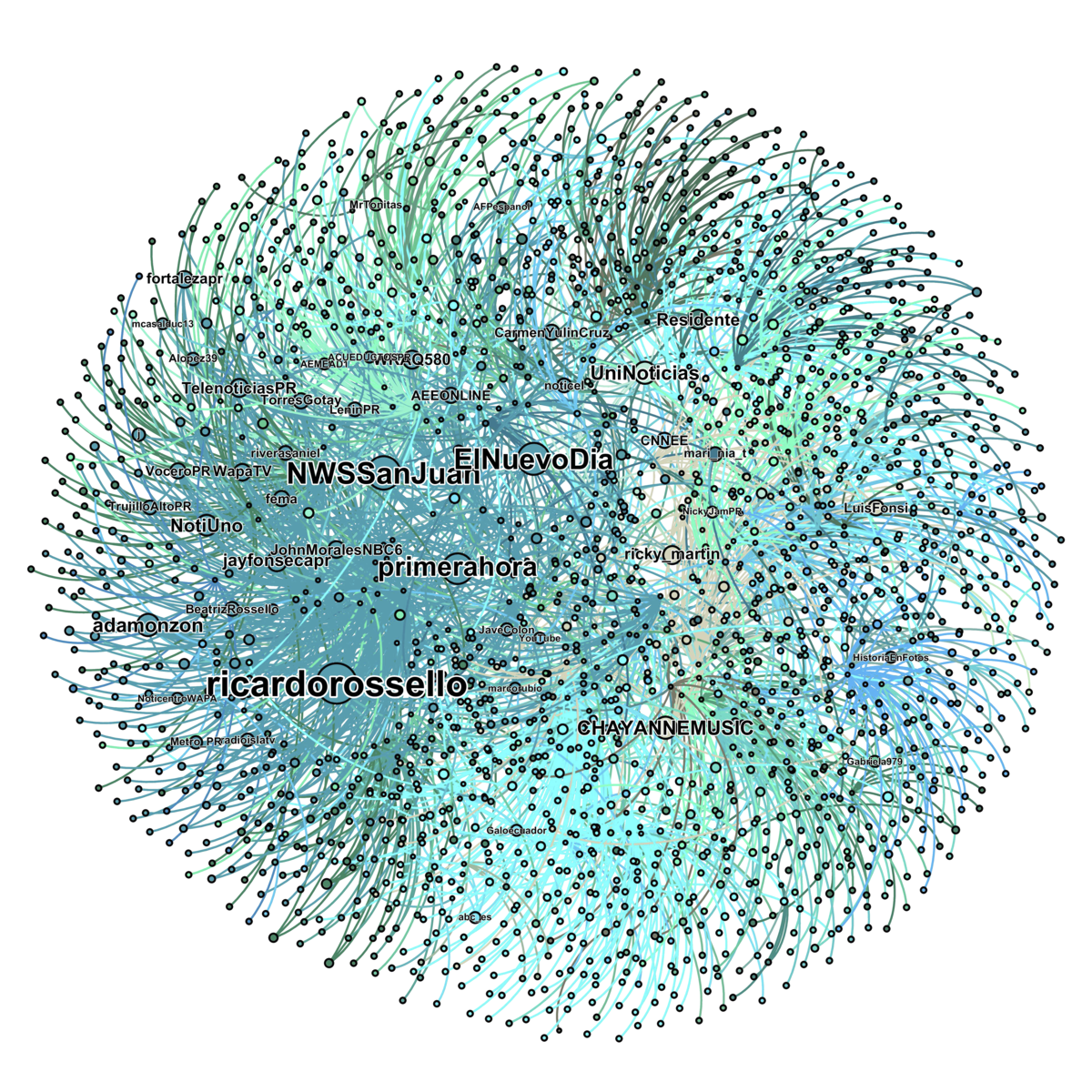}}
\subfigure[\ Before]{\label{fig:networkb}\includegraphics[width=.4\textwidth]{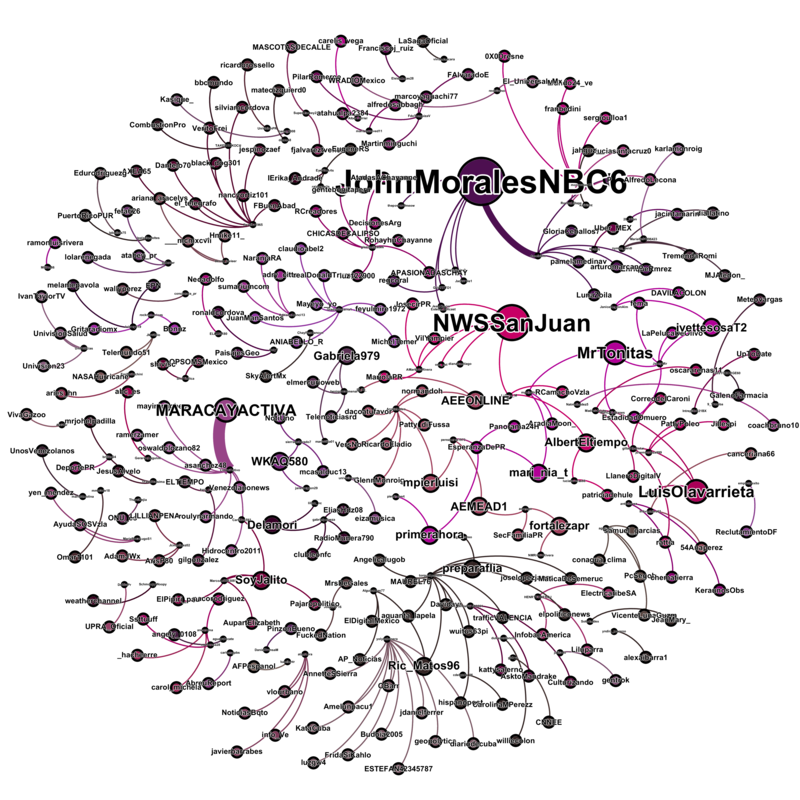}}
\subfigure[\ After]{\label{fig:networkc}\includegraphics[width=.4\textwidth]{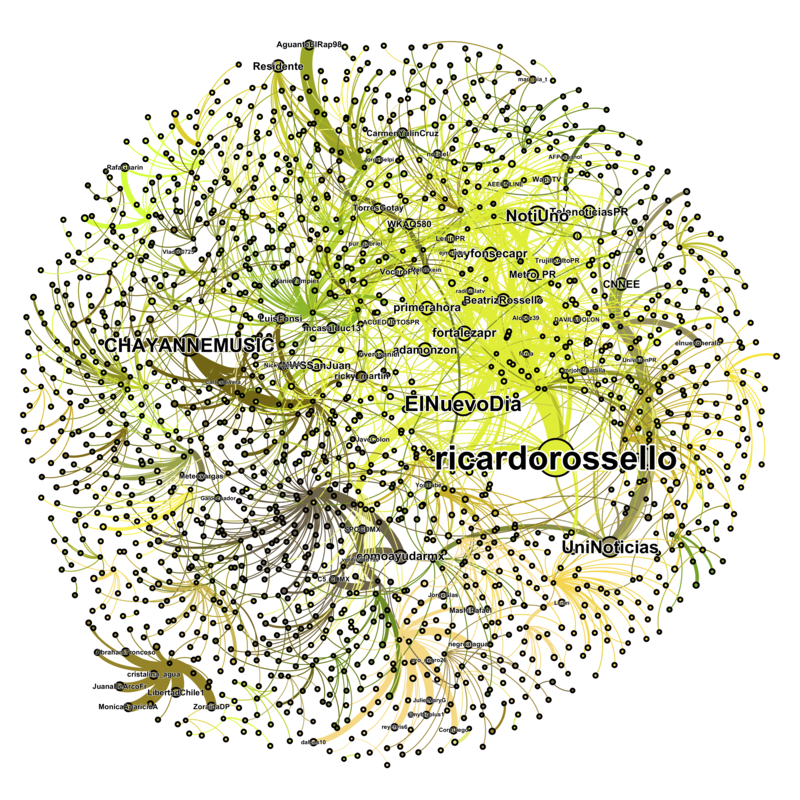}}
\caption{
    The networks of tweets using one or more of the keywords in Tab. \ref{table:keyWords} from users likely to be affected by Hurricane Mar\'ia or to have a direct connection to someone affected: (a) September 16 to October 15; (b) Before hurricane landfall on September 20; and (c) After the hurricane leaves the island.  
  }
\end{figure*}

The majority of nodes in the aggregate network described in Section \ref{sec:methodsNetwork} have in-degree of 0 or 1, while only four have an in-degree greater than 15. This indicates that most of the specified users during this time period were retweeted or mentioned ten or fewer times, where a small minority were retweeted or mentioned more than 150 times. We show full in-degree distribution in Fig. \ref{fig:degreeDist}. This is consistent with the heavy tails usually seen in online social network degree distributions \cite{kumar2010}. Among the fifty highest ranked Twitter accounts by in-degree of this network, sixteen are news outlets, eight are journalists, six are agencies (government and NGO), six are musicians, four are weather stations or meteorologists, and three are citizens.

We display the topology of the communication networks in Figures \ref{fig:networka}, \ref{fig:networkb}, and \ref{fig:networkc}. The size of each node is proportional to the node's in-degree, or the number of tweets tagging or retweeting that user. Communities detected in the network topology by the Louvain clustering algorithm are colored accordingly. Figure \ref{fig:networka} shows the aggregate network for September 16 to October 15 2017. Figures \ref{fig:networkb} and \ref{fig:networkc} show the network during anticipation of the hurricane (September 16--September 19) and during the aftermath (September 22--October 15).

The aggregate network has one major community of more than 300 nodes, which holds the majority of the highest ranked nodes by in-degree. Alongside this community, there are three communities of around 100 nodes. The remainder of the detected communities have fewer than 100 nodes, and most have fewer than 20. The four users with in-degree greater than 15 are all in the largest community. Those are Puerto Rican Governor Ricardo Rossello, one weather station, and two news outlets. The network has a density of $\rho = 6.1\times10^{-4}$ and the average shortest path-length is $\langle \ell \rangle= 1.3$. These measurements indicate that although the network is very sparse, the expected distance between two nodes is small. This is consistent with the heavy-tailed nature of the degree distribution we see in Figure \ref{fig:degreeDist}.  

The anticipation network in Fig. \ref{fig:networkb} has 234 nodes and 261 links, and centers mainly around two accounts: A meteorologist from Florida and a Puerto Rican weather station. Apart from this, the network is made up of many smaller separate connected components between relatively small nodes. The aftermath network has 1,347 nodes and 1,874 links. This subset of the aggregate network strongly resembles the entire aggregate, with the collection of musicians having more prominence, while maintaining a large community of news outlets and weather stations.

\subsection{Divergence}
\label{sec:resultsDivergence}

\begin{figure}[tp!]
  \centering	
    \includegraphics[width=\columnwidth]{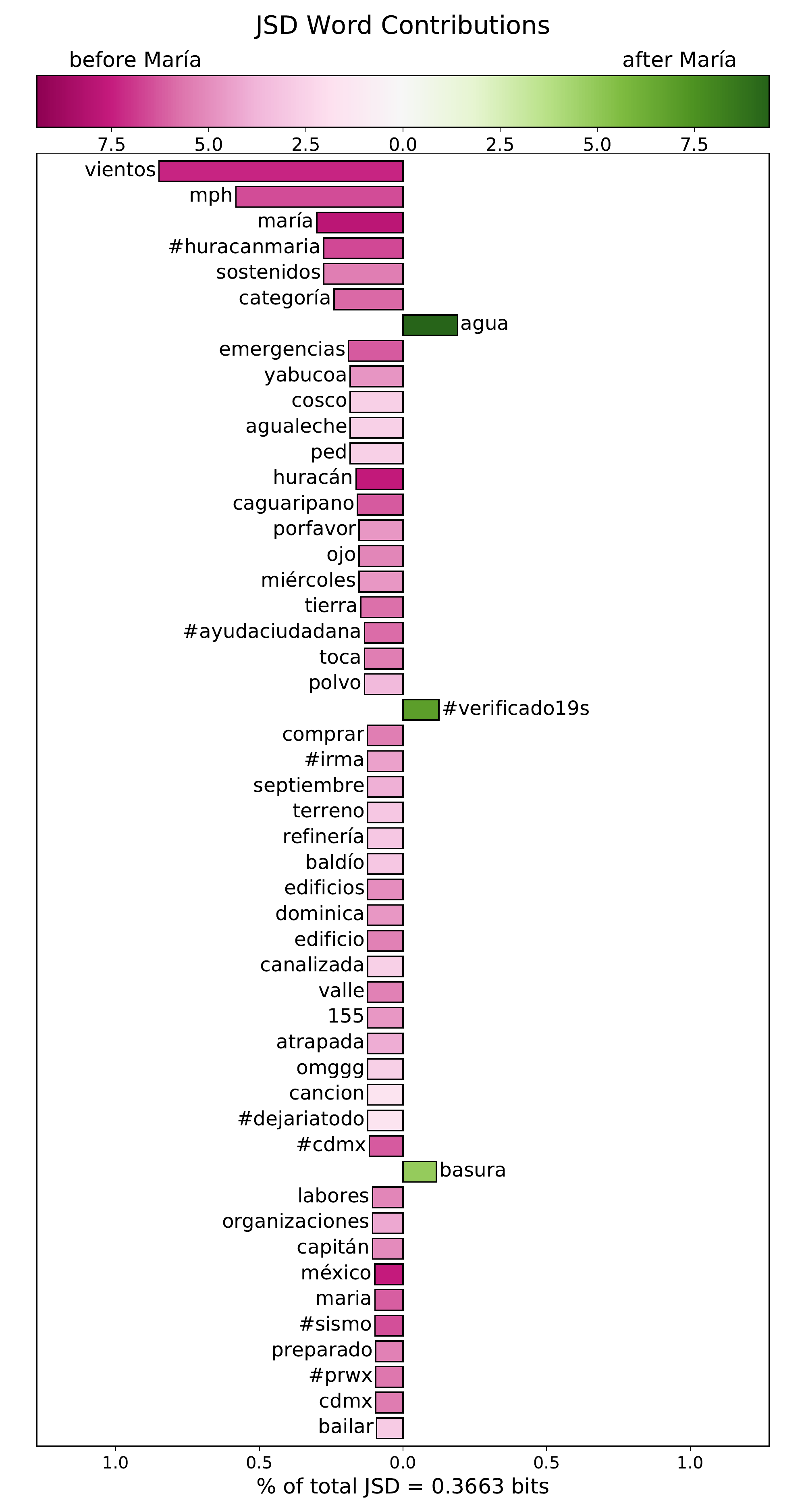}  
  \caption{
    The contributions of individual words to the Jensen-Shannon Divergence between the collection of Tweets before and after the period of landfall for Hurrican Mar\'ia. The length of each bar indicates the contribution of a word, and its direction indicates the corpus in which it was more frequent. The bars are colored according to the Shannon entropy of the Tweets that contained that word, hashtag, or emoji.
  }
  \label{fig:bashift}
\end{figure}

We use Jensen-Shannon divergence wordshifts to demonstrate differences in subsets of our collection of tweets. The length of a bar corresponds to that word's contributions to the Jensen-Shannon divergence between the two corpora, a measurement of the difference in the two word distributions. The bars are colored according to the Shannon entropy of the collection of tweets that use the word, meaning that darker colors indicate a higher diversity of tweet content and colors closer to white indicate more homogeneous content. Bars that are very close to white indicate that the subcollection of tweets using the word are mostly retweets of one specific tweet. In each wordshift, the fifty words with the highest contribution are shown. Emojis are each considered to be one individual word for this analysis.

In Fig. \ref{fig:bashift}, we show the words contributing to the difference between the tweets in anticipation of the hurricane and the tweets from the aftermath. The large majority of the divergence between these two corpora comes from specific words used more in anticipation. Most of these have to do with the severity of the imminent storm. An increase in the frequency of ``agua'' after the hurricane contributes to this divergence, as well as  ``\#verificando19s'', a hashtag used in Mexico to organize a rescue effort in the aftermath of an earthquake on September 19.

\begin{figure*}[tp!]

\centering

\subfigure[]{\label{fig:rcshift}{\includegraphics[width=\columnwidth]{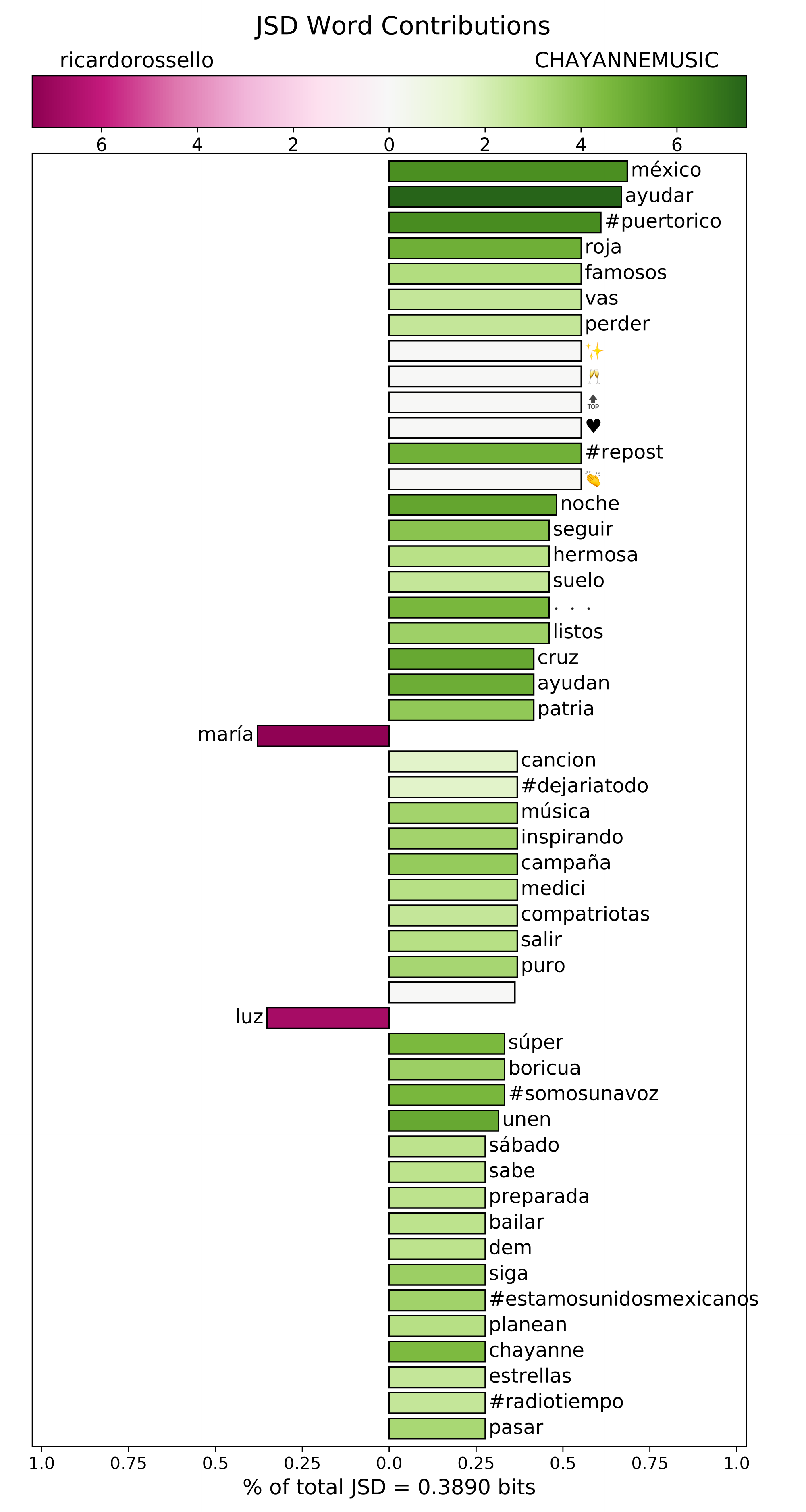}}}
\subfigure[]{\label{fig:rmshift}{\includegraphics[width=\columnwidth]{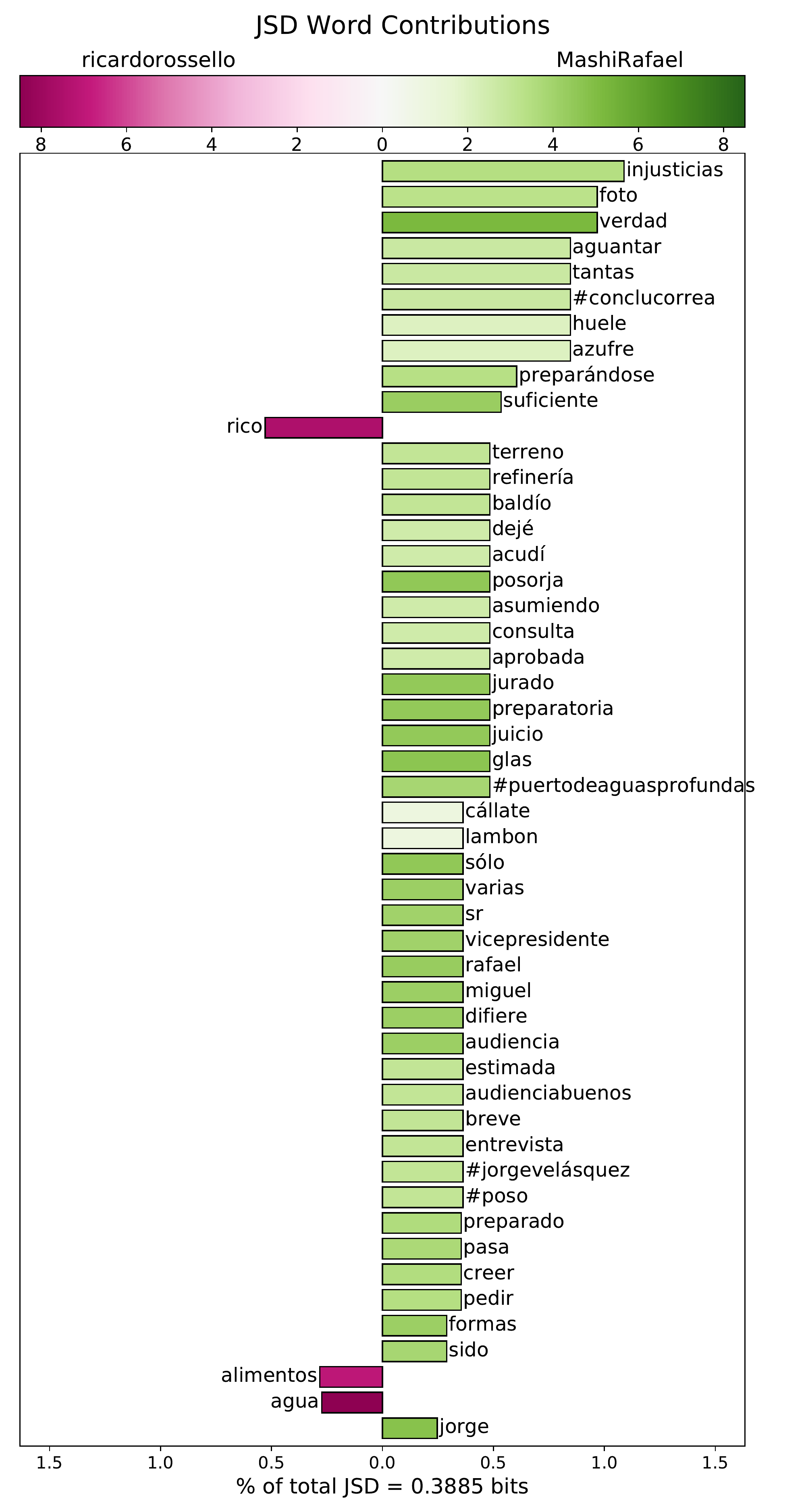}}}

\caption{
     (a) The word-level contribution to the Jensen-Shannon divergence (JSD) between the collection of tweets in the largest and fourth-largest communities. (b) The word-level contribution to the JSD between the collection of tweets in the largest and second-largest communities.
  }
\label{fig:moreShifts}
\end{figure*}

The corpora compared in Fig. \ref{fig:rcshift}'s wordshift are the tweets from the central community containing Governor Rossello's account to those from the fourth-largest community, which contains several famous Puerto Rican musicians such as Chayanne, Luis Fonsi, and Ricky Martin. Although this community is dominated by famous Puerto Ricans, discussion within this community seems to give space to the Mexican earthquake as well, as indicated by the prominence of ``m\'exico'' and the relative infrequency of ``mar\'ia'' and ``luz'', which directly translates to ``light'', but colloquially means ``electricity'' or ``power''.

\subsection{Daily resolution network evolution}
\label{sec:dailyNets}
\begin{figure*}[tp!]

\centering

\includegraphics[width=\textwidth]{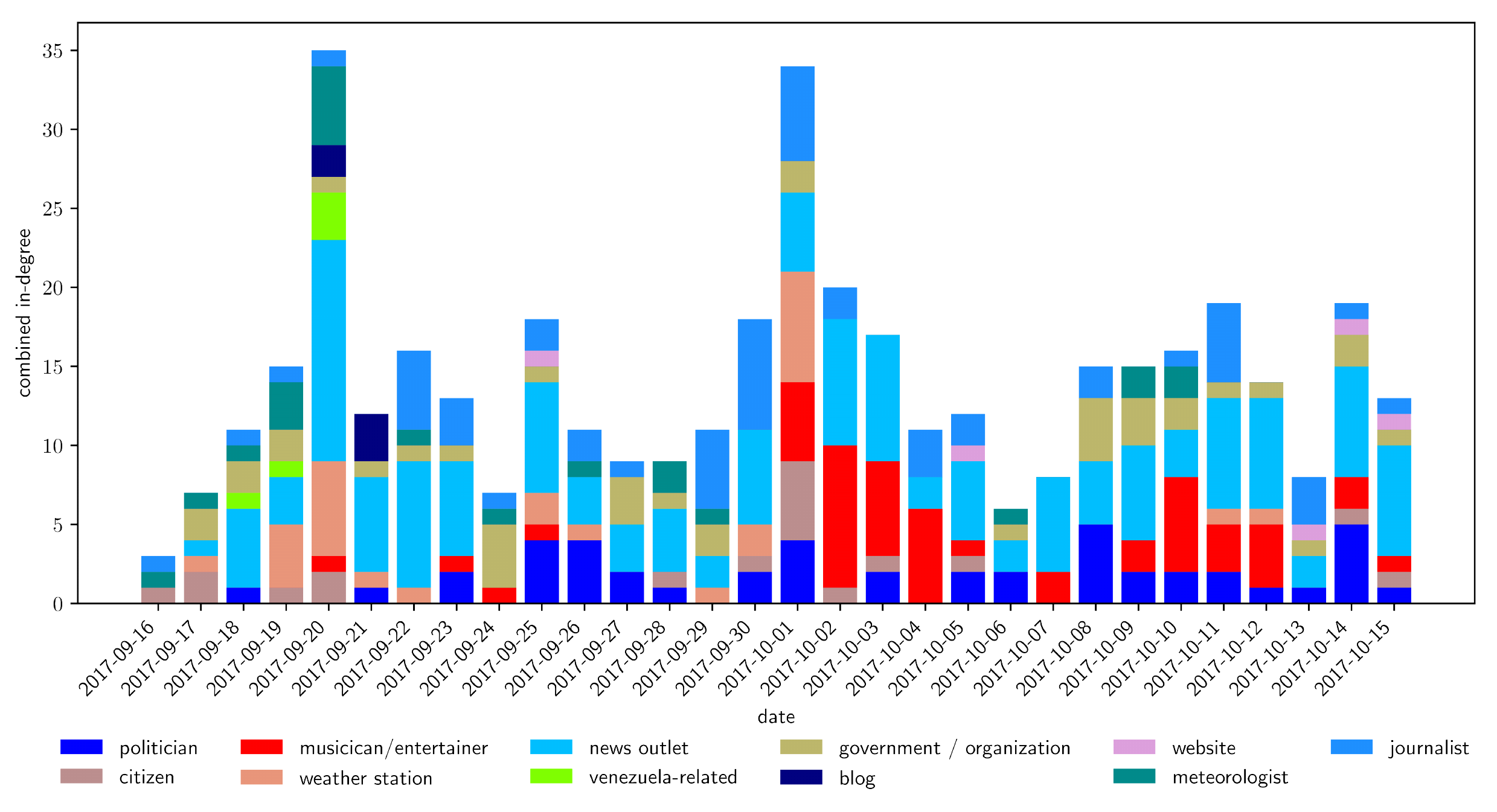}

\caption{
     Stacked bar chart time-series showing the daily in-degrees of nodes in the top-fifty by in-degree of the aggregate network. The nodes are consolidated by manually determined type, shown in the legend. Of these nodes, the news outlets are the most consistently prominent, although politicians and journalists are fairly consistently present in the network as well. We find a spike in activity from musicians starting on October first, and decaying away by the sixth. Musician activity comes back for an interval of four days starting on October 9th, but this is all due to one single account, Residente. This can be seen in Figure \ref{fig:timeSeries-ind} in the supplementary materials, where we show the same timeseries separated by individual accounts.  
  }
\label{fig:timeSeries}
\end{figure*}

We separate the aggregate network into its daily subnetworks, and look at the top fifty users in the aggregate network at the daily resolution. Fig. \ref{fig:timeSeries} illustrates the change of in-degree over time of these nodes in the daily networks, binned by user type. We see that news outlets are a relatively consistent player in the conversation, whereas weather stations and meteorologists were only prominent before the hurricane and on the first day of October. Government and other aid organizations tend to take up a small portion of the conversation, becoming major players somewhat suddenly on September 24 and October 8--9.

The same figure with every individual user colored separately can be found in the Appendix. In examining individual users with time, we find that in the days surrounding the landfall of the hurricane, the tweets from the meteorology station @NWSSanJuan are heavily propagated. A few days after the end of the rain, Governor Rossello becomes a prominent voice in the network on most days from then on. During the first few days of October, Puerto Rican singers Chayanne, Ricky Martin, and Luis Fonsi occupy much of the conversation space. Later on, starting on October 10, Puerto Rican rapper Residente becomes a major node in the network, but he appears to mostly connect with smaller nodes, and not as much with other major figures.

\section{Discussion}
\label{sec:discussion}

\subsection{Global conversations}

The happiness levels of each language within the body of tweets about the hurricane is very stable, especially when compared to the same window of time for the collection of all English tweets (http://hedonometer.org). There are a handful of incidents where the sentiment trajectories of the two languages track one another, as well as a set of instances where they spike independently. This can be reconciled with the varying scales of impact that different events may have. 

An independent spike in Spanish sentiment on September 19 is an intriguing event, as it conveys an important realization about how Spanish speakers spoke about the two major natural disaster events that were happening or were imminent. It can be seen from the wordshift from that day that Spanish speakers are discussing the earthquake in Mexico along with the incoming hurricane. The influential words, however, seem to indicate an emphasis on hope and the inspiring action of helpers. Words that stand out here are ``h\'eroes'', ``dios'', and ``ni\~nos''. It appears that there is a focus among Spanish tweeters on religious figures and those demonstrating heroism in the wake of the disaster, including the young boys who were delivering supplies on mountain bikes to the first responders \cite{fountain2017packs}.

Our findings point to other methods of investigating the true sources of difference and similarity in the discourse between communities who speak different languages. For instance, one might apply wordshift methods to investigate dips by comparing them to the corpora from previous days. One might also wish to examine information theoretic trends in each of the languages, such as normalized entropy.

\subsection{Local conversations}

In a previous work we have found that, in response to a major natural crisis event, average users with small followings increased their social media activity more than influential accounts with large followings \cite{niles2018average}. While it is challenging to identify those responsible for the propagation of information throughout the Twitter network, our work here begins to address the natural follow on question: Where is the information originating? The consistent dominance of local journalists, news outlets, and politicians in the communication network demonstrates the accounts people turned to were those whose audience is exclusively Puerto Rico on any given day of the year. 

Our results suggest that crisis information is spread throughout the network by average individuals, the first of which get such information from ``local'' Twitter users, such as the politicians and journalists from Puerto Rico. Meanwhile, major celebrities seem mostly removed from a conversation centered on local actors and average affected individuals, touching the Puerto Rico crisis along with a variety of issues in Latin America.

There are several important limitations to acknowledge surrounding this work and its possible implications. The data Twitter provides via its API distorts the network measurements by artificially rewiring retweet chains into stars centered on the original author of the tweet. Network measurements are also altered by the use of only a 10\% sample of tweets. Our methods of identifying tweets from Puerto Ricans and those with close connections surely introduced some error, despite being our best resort due to the very low frequency of geolocated tweets. Another compounding problem is that most people of interest had no cellular phone reception for the majority of the studied time interval. We note, however, that this means our network presents a lower bound on representation of regular Puerto Ricans and accounts with a mostly Puerto Rican audience. Our main finding, that the most consistent providers of information were those with more moderate localized audiences, is a conservative one. The true nature of the situation may very well be stronger.

\section{Conclusion}
\label{sec:conclusion}

Time and time again people prove incredibly resilient under conditions of extreme diversity. In the case of Hurricane Mar\'ia, Puerto Ricans took care of themselves and their families while facing a barrage of extreme weather, from flooding to heat waves. Further, they searched for ways to communicate with those outside their immediate presence, and found such ways on the sides of highways across the island. 

The results we present hold implications for disaster relief, and information dissemination during major crisis events. For the most part we find that the best accounts to provide important information tend to be local figures: Journalists, news outlets, politicians. While we did see points in time during which celebrities became central, they did not remain reliably central for any major span of time within the studied interval, which included anticipation, event, and aftermath. Meanwhile, content from global tweets indicated the a focus on hope and resilience from Spanish speaking users.

In any natural crisis situation, there is almost never a shortage of helpers. The emergence of massive cooperating teams during times of need is a signature of humanity. Lack of information, or explicit blockages that keep information from those who need it, however, can render this help not useful to the overwhelming masses of crisis victims. It is clear that the transfer of information about where and when to find help is as important as the help itself. We hope our results can inform institutions regarding how to use Twitter to best share information so that it most efficiently propagates through the network, getting to as many people as possible, as quickly as possible, and remaining uncorrupted in content.

\vspace{1em}
\acknowledgments
The authors are grateful for support and helpful feedback from the members of the Computational Story Lab and the graduate and undergraduate advisees of Dr. Meredith Niles. CMD and PSD were supported in part by a gift from MassMutual Insurance

\bibliography{\filenamebase.bib}

\clearpage

\newwrite\tempfile
\immediate\openout\tempfile=startsupp.txt
\immediate\write\tempfile{\thepage}
\immediate\closeout\tempfile

\setcounter{page}{1}
\renewcommand{\thepage}{S\arabic{page}}
\renewcommand{\thefigure}{S\arabic{figure}}
\renewcommand{\thetable}{S\arabic{table}}
\setcounter{figure}{0}
\setcounter{table}{0}
\begin{figure*}[tp!]

\centering

\includegraphics[width=\textwidth]{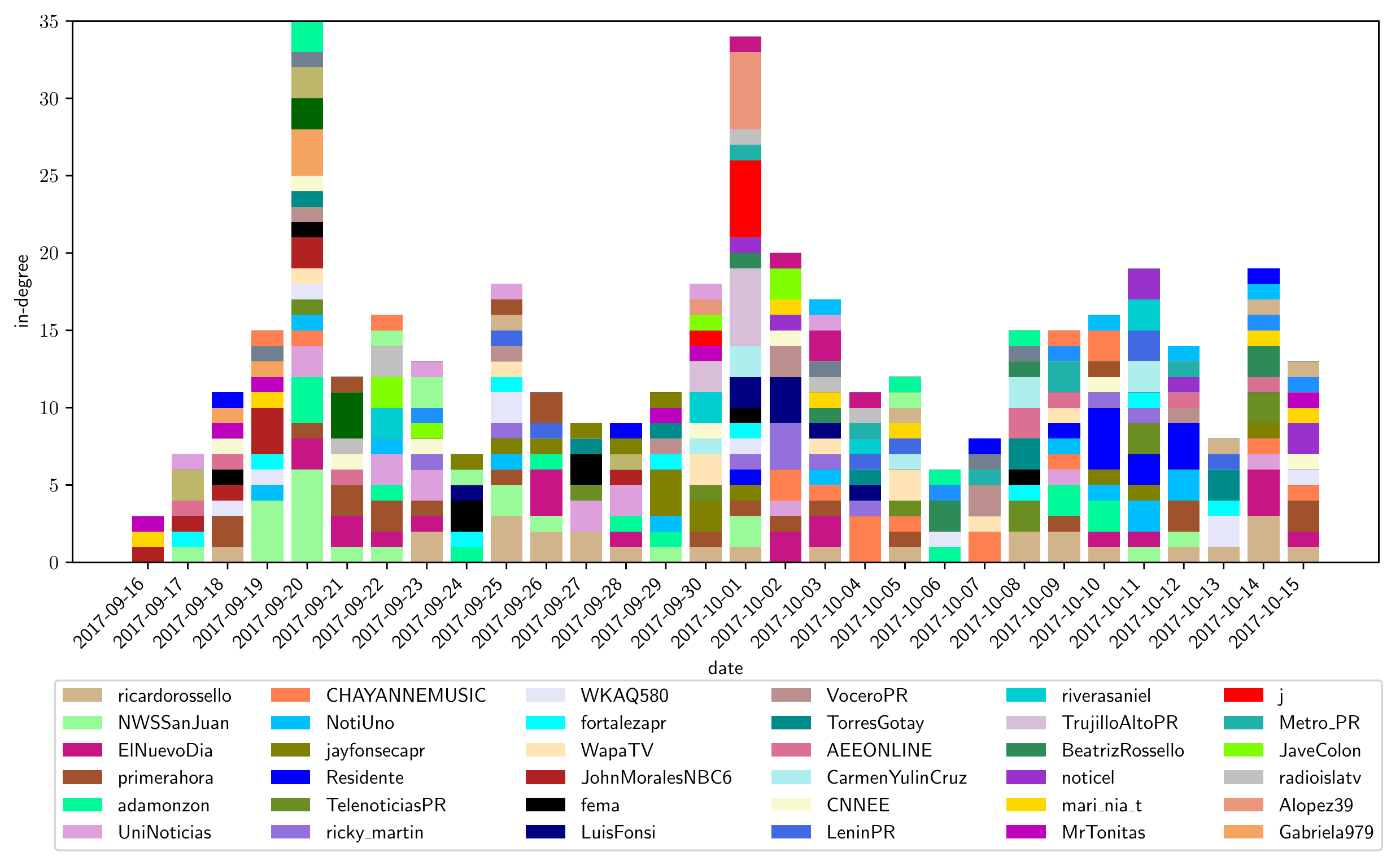}

\caption{Stacked bar chart time-series showing the daily in-degrees of nodes in the top-fifty by in-degree of the aggregate network. Here, the individual nodes are broken out of their categories shown in Fig. \ref{fig:timeSeries} and labeled separately to show individual contribution.
}

\label{fig:timeSeries-ind}
\end{figure*}

\end{document}